\documentclass[fleqn]{annalen}
\usepackage{graphics}
\begin{document}
\newcommand{\volume}{8}              
\newcommand{\xyear}{1999}            
\newcommand{\issue}{5}               
\newcommand{\recdate}{29 July 1999}  
\newcommand{\revdate}{dd.mm.yyyy}    
\newcommand{\revnum}{0}              
\newcommand{\accdate}{dd.mm.yyyy}    
\newcommand{\coeditor}{ue}           
\newcommand{\firstpage}{507}         
\newcommand{\lastpage}{510}          
\setcounter{page}{\firstpage}        
\newcommand{\keywords}{disordered solids, quantum magnets, rare regions}
\newcommand{\PACS}{75.20.Hr, 75.10.Jm, 75.30.Kz}
\newcommand{\shorttitle}{R. Narayanan et al., Rare regions in quantum magnets}
\title{Rare regions and annealed disorder in quantum phase transitions}
\author{R. Narayanan$^{1}$, T. Vojta$^{1,2}$, D. Belitz$^{1}$ and
    T. R. Kirkpatrick$^{3}$} 
\newcommand{\address}
{$^1$Dept. of Physics and Materials Science Institute,
                                      University of Oregon, Eugene, OR 97403\\
$^2$Institut f{\"u}r Physik, TU Chemnitz, D-09107 Chemnitz, FRG\\
$^3$IPST, and Department of Physics,
     University of Maryland, College Park, MD 20742\\}
\newcommand{\email}{\tt rajesh@oregon.uoregon.edu} 
\maketitle
\def\tr{{\rm tr}\,}
\def\Tr{{\rm Tr}\,}
\def\sgn{{\rm sgn\,}}
\def\b{\bibitem}
\def\boldphi{\mbox{\boldmath $\phi$}}
\def\boldvarphi{\mbox{\boldmath $\varphi$}}
\begin{abstract}
The Griffiths region that is due to rare regions and the resulting local
moments in disordered itinerant 
quantum magnets, and its influence on the critical behavior, is considered 
within the framework of an effective field theory. It is shown that the
local moments can be described in terms of static, annealed disorder,
and the physical consequences of this description are discussed.
\end{abstract}

\section{Introduction}
\label{sec:I}

In systems with impurities, one distinguishes
between two types of disorder: The impurities may be in equilibrium with
the other degrees of freedom (``annealed'' disorder), or
`frozen in' (``quenched'' disorder). 
In the former case the free energy is obtained in
terms of the disorder averaged partition function, while
in the latter usually the free energy is a self-averaging 
quantity \cite{Grinstein}. In magnets, quenched disorder
reduces the critical temperature from
its clean value $T_c^0$ to $T_c<T_c^0$, and the magnetization $M$ is a
nonanalytic function of the external magnetic field $B$ everywhere in the
``Griffiths region'' $T_c<T<T_c^0$ \cite{Griffiths}. This is a result of 
the existence of rare regions that are devoid of any impurities and lead to
the formation of locally ordered regions, or local moments, even though
the system as a whole does not display global order. Since large rare
regions are exponentially unlikely, this is a weak effect in generic
classical systems; the singularity in the free energy is only an essential
one.

\begin{figure}
\centerline{\resizebox{9.0cm}{4.5cm}{\includegraphics{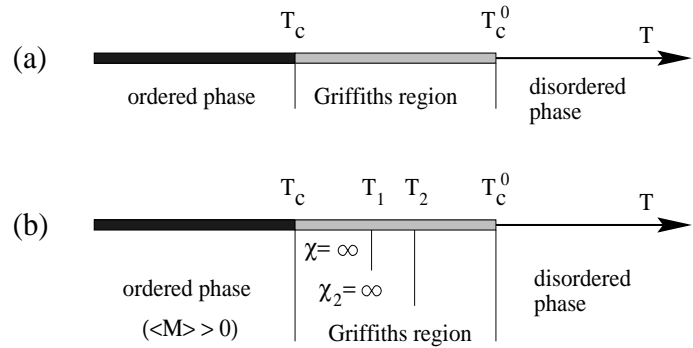}}}
\caption{Schematic phase diagram for (a) a classical ferromagnet with 
         uncorrelated disorder, and (b) the McCoy-Wu model.}
\label{fig:1}
\end{figure}

A stronger effect of the same type is encountered in classical models
with correlated disorder. McCoy and Wu \cite{McCoyWu} studied a
$2$-$d$ Ising model with identical bonds in $y$-direction, and bonds in
$x$-direction that are random, but identical within each column. Here
the rare regions lead to a hierarchy of temperatures $T_n$ ($n=1,2,\ldots$)
between $T_c$ and $T_c^0$, with the disorder averaged $n^{\rm th}$ order
susceptibility, $\chi_n = \langle\partial^n M/\partial B^n\rangle$,
diverging for all $T<T_n$. Even the susceptibility proper,
$\chi\equiv\chi_1$, diverges for $T_c<T<T_1$, even though 
$\langle M\rangle$ becomes nonzero only for $T<T_c$, see 
Fig.\ \ref{fig:1}. The reason for this are atypical 
fluctuations in the distribution of $\chi$ which dominate the mean \cite{DSF}.
A generalization to $d$ dimensions, with the disorder
correlated in one of them, is very similar to models of {\em quantum}
magnets in $d-1$ spatial dimensions with {\em uncorrelated} disorder, since
imaginary time plays the role of the correlated direction.
One therefore expects strong effects of rare
regions on generic quantum magnets. This is borne out by studies of
quantum spin system in $d=1$ \cite{DSF}. However, little it known
about the Griffiths phase and its influence on the critical properties
in quantum systems in $d>1$.

Motivated by similar considerations, we have recently considered rare 
regions in quantum magnets in $d>1$\cite{us_rr}. Previous theories for 
itinerant antiferromagnets \cite{us_afm} and ferromagnets \cite{us_fm}
had found power-law critical behavior. It was found in Ref.\ \cite{us_rr}
that rare regions destroy the
conventional fixed point in the former case, but not in the latter.
In this paper we discuss the basic physics behind these findings,
in particular the notion of local moments as static annealed disorder,
and the relation between this physics and some important technical
points that arise in the derivation of a model that describes local moments.

\section{Itinerant quantum antiferromagnets}
\label{sec:II}

We consider the following action for
an itinerant quantum antiferromagnet\cite{us_rr},
\begin{equation}
\hskip -10pt S = \int dx\,dy\ {\boldphi}(x)\,\left[
      \Gamma^{(0)}(x-y) + \delta(x-y)\,\delta t({\bf x})\right]\,{\boldphi}(y)
    + u\int dx\ \left({\boldphi}(x)\cdot{\boldphi}(x)\right)^2.
\label{eq:1a}
\end{equation}
Here $x\equiv ({\bf x},\tau)$ comprises position ${\bf x}$ and imaginary
time $\tau$, $\int dx \equiv \int d{\bf x}\int_{0}^{1/T}d\tau$, and we
put $\hbar = k_{\rm B} = 1$. 
The Fourier transform of $\Gamma^{(0)}$ is (omitting constant prefactors)
\begin{equation}
\Gamma^{(0)}({\bf q},\omega_n) = (t + {\bf q}^2 + \vert\omega_n\vert)/2\quad.
\label{eq:1c}
\end{equation}
Here $t$ denotes the distance from the critical point, and $\delta t$ in
Eq.\ (\ref{eq:1a}) is a random function of position.

In analogy to the treatment of classical random magnets by Dotsenko et al.
\cite{Dotsenko}, we now formally consider inhomogeneous saddle-point
solutions of our field theory in the disordered phase ($t>0$). Such solutions
exist since the disorder allows for order parameter configurations
that are nonzero on `islands' where
$t+\delta t({\bf x})<0$, and they describe a Griffiths phase. 
Let there be $N$ such islands. Since they are far 
apart, we actually have $2^N$ almost degenerate saddle-point solutions,
$\phi_{\rm sp}^{(a)}$ $(a=1,\ldots,2^N$),
that are obtained by flipping the local magnetizations on the individual
islands.
The partition function can be calculated by expanding about a particular
saddle point, say, $\phi_{\rm sp}^{(1)}$:
$Z = \int D[\varphi]\ \exp[-S(\phi_{\rm sp}^{(1)} + \varphi)]$.
This is exact as long as the functional integral is extended
over all fluctuations $\varphi$ of the order parameter field. In practice,
however, one can perform the integral only perturbatively, taking into
account only small fluctuations. Simple energy and statistics considerations
show that, in the thermodynamic limit, almost all of the nearly degenerate 
saddle point configurations are separated from one another by macroscopic
energy barriers, so any perturbative evaluation of the integral 
defining $Z$ will miss a macroscopic number ($2^N-1$) of contributions
to $Z$ that are equally important as the one obtained by expanding about
$\phi_{\rm sp}^{(1)}$. However, it also means that, as long as we confine
ourselves to a perturbative evaluation of the functional integral, we can
simply add up these contributions:
\begin{equation}
Z \approx \sum_{a=1}^{2^N} \int_{<} D[\varphi]\ 
                                  e^{-S[\phi_{\rm sp}^{(a)} + \varphi]}
   = \int D[\phi_{\rm sp}]\ P[\phi_{\rm sp}] \int_{<} D[\varphi]\ 
                                  e^{-S[\phi_{\rm sp} + \varphi]}\quad.
\label{eq:3b}
\end{equation}
Here $\int_{<}$ indicates an integration over small fluctuations only, and
we have replaced the sum over
saddle points by an integration over a suitable distribution 
$P[\phi_{\rm sp}]$. 

The local moments $\phi_{\rm sp}({\bf x})$ are
described by a random function of position, since they are ultimately
determined by the random $\delta t({\bf x})$.
However, since they are generated by the electron system
itself, in response to the random potential, they are in
equilibrium with the rest of the electronic degrees of freedom. Therefore,
the partition function is averaged over
the saddle-point configurations, which hence represent static,
annealed disorder.

We now perform the averages over the disorder. In a cumulant expansion, 
any reasonable distribution $P[\phi_{\rm sp}]$ will in particular produce 
a term that is produced by a Gaussian distribution, viz.
$w\int d{\bf x}\int d\tau\,d\tau'\ (\varphi({\bf x},\tau))^2\,
(\varphi({\bf x},\tau'))^2$ with some coupling constant $w$. 
The quenched disorder we handle
by means of the replica trick \cite{Grinstein}. In a Landau expansion,
and with $\Gamma^{(0)}$ from Eq.\ (\ref{eq:1c}),
we obtain the following effective action:
\begin{eqnarray}
S_{\rm eff}&=&\sum_{\alpha} \int dx\,dy\ \boldvarphi^{\alpha}(x)\,
                   \Gamma^{(0)}(x,y)\,\boldvarphi^{\alpha}(y)
               + u\sum_{\alpha}\int dx\ \left(\boldvarphi^{\alpha}(x)\cdot
                                    \boldvarphi^{\alpha}(x)\right)^2
\nonumber\\
&&- \sum_{\alpha,\beta}(\Delta + w\,\delta_{\alpha\beta})\int dx\,dy\,
  \delta({\bf x}-{\bf y})\,
  \left(\boldvarphi^{\alpha}(x)\right)^2\,
   \left(\boldvarphi^{\beta}(y)\right)^2
 + O(\boldvarphi^6)\quad.
\label{eq:4}
\end{eqnarray}
Here $\Delta$ is the variance of the $\delta t$-distribution,
and $\alpha$ and $\beta$ are replica indices.

An RG analysis of $S_{\rm eff}$
reveals that the $w$-dependent terms in the flow
equations contain factors of $1/T$, and hence do not exist at $T=0$.
This can be understood by means of a simple zero-dimensional toy model
that couples Gaussian disorder $v$, with variance $w$, to an order parameter
$m$. Integrating out the disorder yields
\begin{equation}
\int dv\ e^{-v^2/w - 2vm/T} \propto e^{wm^2/T^2}
\label{eq:5}
\end{equation}
This leads to a term in the free energy that is proportional to $1/T$.
The reason for the occurrence of these unphysical terms is the unboundedness
of the disorder distribution: Since annealed disorder allows for full
equilibration, the system can always lower its free energy by choosing
deeper and deeper potential wells. This unphysical feature can be easily
cured by using a distribution that is bounded below. A mathematically and
physically equivalent procedure is to make the variance of the unbounded
distribution a linear function of temperature. We therefore write
$w = {\bar w}T$, and consider ${\bar w}$ as the coupling constant that
represents the presence of local moments. We note that this is a fundamental
feature of annealed disorder. 
The fact that the variance of any annealed disorder
distribution effectively vanishes as $T\rightarrow 0$ has important
consequences for all problems involving local moments, since it implies
that the mass that some modes acquire due to the local moments vanishes
at zero temperature. The massive modes therefore do not simply drop out
of the problem, as one might naively expect.

A standard one-loop RG analysis shows that the
critical fixed point discussed in Ref.\ \cite{us_afm} is unstable with
respect to the new coupling constant ${\bar w}$ that represents the rare 
regions, and that one finds runaway flow in all
physical regions of parameter space. The interpretation of this runaway flow
is currently not clear. Possibilities include, the absence of a phase
transition, or, more likely, the existence of a fixed point that is not
accessible by means of a perturbative analysis of our effective action.

An analysis of itinerant quantum ferromagnets 
can be performed along the
same lines\cite{us_rr}. In this case, 
an effective long-range
interaction between the order parameter fields suppresses all
fluctuations, including those due to the
rare regions. As a result, the
critical behavior of itinerant quantum ferromagnets as determined in
Ref.\ \cite{us_fm} is not affected by rare regions, in sharp contrast to
the antiferromagnetic case.

Finally, we point out that the treatment of rare regions sketched above,
and the physical consequences of the properties of annealed disorder we
have discussed, are very general and have important consequences for
other problems where local moments are important, e.g., for
metal-insulator transitions. Results for such problems will be
reported elsewhere.

\vspace*{0.25cm} \baselineskip=10pt{\small \noindent
We thank F. Evers and J. Toner for discussions. This work was supported by 
the NSF (DMR-98-70597 and DMR--99--75259), and by the DFG  (SFB 393/C2).}

\end{document}